\documentclass[10pt,amsmath,amssymb,a4paper,twocolumn,pra]{revtex4}
\usepackage{theorem} 
\usepackage{graphicx}

\newcommand{\Hi}{\mathcal{H}} 
 
\newcommand{\Tr}{\mathrm{Tr}}

\newcommand{\lv}{\left \vert}
\newcommand{\rv}{\right \vert}
\newcommand{\la}{\left \langle}
\newcommand{\ra}{\right \rangle}
\newcommand{\ket}[1]{\lv #1 \ra}
\newcommand{\bra}[1]{\la #1 \rv}

\theorembodyfont{\rmfamily}
\newtheorem{Theorem}{{Theorem} }

\newtheorem{Lemma}{{Lemma}}
\newtheorem{Proof}{{Proof}}

\begin{document}

\title{Local copying of orthogonal maximally entangled states \\ and its relationship to local 
discrimination\footnote{The material in this paper was presented on November 15, 2004 at 
Cambridge workshop on Quantum Statistics - Quantum Measurements, 
Estimation and Related Topics, Cambridge, UK.}} 
\author{Masaki Owari$^{1,2}$, Masahito Hayashi$^{1}$} 
\address{$^1${\it ERATO Quantum Computation and Information Project, JST, Tokyo 113-0033, Japan}\\ $^2${\it Department of Physics, The University of Tokyo, Tokyo 113-0033, Japan}}

\begin{abstract}
In the quantum system, perfect copying is impossible without prior 
knowledge.
But, perfect copying is possible, if it is known that unknown states to 
be copied is contained by the set of orthogonal states, which is called the 
copied set.
However, if our operation is limited to local operations and classical 
communications, this problem is not trivial.
Recently, F. Anselmi, A. Chefles and M.B. Plenio constructed theory of local 
copying when the copied set consists of maximally entangled states.
They also classified the copied set when it consists of two orthogonal states 
({\it New. J. Phys.} {\bf 6}, 164 (2004)).
In this paper, we  completely classify the copied set of local copying of the 
maximally entangled states in the prime dimensional system. 
That is, we prove that, in the prime dimensional system, the set of locally 
copiable maximally entangled states is equivalent to the set of simultaneously
Schmidt decomposable canonical form Bell states.
As a result, we conclude that local copying of maximally entangled states is
much more difficult than local discrimination at least in prime-dimensional 
local systems. 
\end{abstract}

\maketitle

\section{introduction}
Both cloning and entanglement are main features of quantum information
and have been treated as central topics in this field.
Many researches have been done for investigation for these two topics
and many remarkable results have been derived.
On the other hand, since cloning and entanglement play central roles
in quantum information, the relationship
between these two issues is crucially important for deeper understanding
of quantum information processings.
However, not so many works have been
done in the above direction except a few paper \cite{cloning-entanglement} \cite{tele-cloning}. 
In this paper, we treat the LOCC copying as one of such important
problems lying between these two issues.

Many researchers treated
approximated cloning, for example, universal cloning \cite{universal
cloning}, asymmetric cloning \cite{asymmetric cloning},
tele-cloning \cite{tele-cloning}.
This is because the perfect cloning, {\it i.e.}, copying, is impossible without prior knowledge \cite{no-cloning}.
That is, the possibility of copying depends on the prior knowledge.
If we know that
the unknown state to be copied is contained by the set of orthogonal states, 
which is called the copied set, we can copied the given state.
However, if the copied system has an entangled structure, and if our operation 
is restricted to local operations and classical communications (LOCCs)\cite{LOCC}, we
cannot necessarily copy the given quantum state with the above orthogonal assumption,
perfectly.
Thus, it is interesting from both viewpoints of entanglement theory and cloning 
theory to extend this problem to the bipartite entangled setting.

Recently, F. Anselmi, A. Chefles and M. B. Plenio \cite{ACP} focused on the perfect copying with the bipartite system under the following assumption;
\begin{enumerate}
\item Our operation is restricted to LOCC.
\item It is known that the unknown state to be copied is contained by the set of orthogonal entangled states, (the copied set).
\item An entangled state with the same size is shared.
\end{enumerate}
They called this problem local copying, and their main point is what copied set enable us to locally copy the given state.
In the following, for simplicity,
we say the set is locally copiable
if local copying is possible
with the prior knowledge to which
the given state belongs.
As is explained in Theorem \ref{nscondition},
they showed that the possibility of local copying can be reduced to the simultaneous transformation of unitary operators.
That is, they derived a necessary and sufficient condition for a locally copiable set.
Moreover, they pointed out
an essential relation between local copying and entanglement catalysis, which is a famous open problem 
in the field of entanglement theory.
This relation indicates
that local copying is very attractive as a problem which possesses specific many features for quantum information.

In this paper, in the first step, we give a typical example of a locally copiable 
set consisting of $D$ maximally entangled states in a $D^2$-dimensional 
bipartite system.
Then, conversely, we prove that
any locally copiable set of maximally entangled states can be regarded as a 
subset of the above typical example if the dimension $D$ of the local space is 
prime.
That is, if we choose a suitable basis
for a given locally copiable set of maximally entangled states, 
it is a  subset of the above typical example.
Finally, we discuss the relationship between local copying and local
distinguishability \cite{WSHV00, VSPM01, CY02, C04, F04}, which is also an 
important topic in relation to entanglement theory.

As our conclusion,
we obtain the following four important results with prime-dimensional local spaces;
\begin{enumerate}
\item The maximum size of the locally copiable set of maximally entangled states is $D$.
\item Any copiable set of maximally entangled states must be represented as the set of powers of Pauli's $Z$ operator.
\item Local copying is more difficult than simultaneous Schmidt decomposition.
\item Local copying is more difficult than local discrimination.
\end{enumerate}

This paper is organized as follows.
In section \ref{preliminary}, we review
a necessary and sufficient condition for a locally copiable set as the preparation of our analysis, 
which is the main result of the paper \cite{ACP}.
In section \ref{local copying}, we give an example of a locally copiable set of $D$ maximally entangled states, 
and then, prove that, in a prime-dimensional local system, the above example is the only case where local copying 
is possible.
In section \ref{relation}, we discuss the relation among local copying,
simultaneous Schmidt decomposition and LOCC discrimination.  
Finally, we summarize
our result in section \ref{summary}.

\section{The preliminary  (review of \cite{ACP})} \label{preliminary}
The problem of local copying can be arranged as the follows.
We assume two players at a long distance, {\itshape e.g.},  
Alice and Bob in this protocol. 
They have two quantum systems $\Hi _A$ and $\Hi _B$ each of which is also composed
 by the same  {\itshape two} $D$-dimensional systems, 
{\itshape i.e.}, the systems $\Hi _A$ and $\Hi _B$
are described
by $\Hi _A = \Hi _1 \otimes \Hi _3$, $\Hi _B= \Hi _2 \otimes \Hi _4$.
In our problem, they try to copy an unknown state $\ket{\Psi}$ on the initial
system 
$\Hi _1 \otimes \Hi _2$ to the target system $\Hi _3 \otimes \Hi _4$
with the prior knowledge that $\ket{\Psi}$ belongs to the copied set
$\{ \ket{\Psi _j } \}_{j=0}^{N-1}$.
Moreover,  we assume that they implement copying only by LOCC between them.
Since LOCC operations do not increase the entanglement of whole states,
they can copy no entangled state by LOCC without any entanglement resource.
Thus, we also assume that they share a blank entangled state $\ket{b}$ in target
systems $\Hi _3 \otimes \Hi _4$.
Therefore, a set of states $\{ \ket{\Psi _j } \}_{j=0}^{N-1}$ 
is called locally copiable with a blank state $\ket{b ^{3,4}} \in \Hi _3 \otimes 
\Hi _4$, 
if and only if there exists
a LOCC operation $\Lambda$ on $\Hi _A \otimes \Hi _B$
which satisfies the following condition for all $ j = 0, \cdots , N-1$: 
 \begin{eqnarray*}
& & \Lambda  (\ket{\Psi _j^{12}} \otimes \ket{b^{34}} \bra{\Psi _j^{12}} \otimes
 \bra{b^{34}} ) \\
 & = & \ket{\Psi _j^{12}} \otimes \ket{\Psi _j^{34}} 
 \bra{\Psi _j^{12}} \otimes \bra{\Psi _j^{34}},
\end{eqnarray*}
 where we treat $\Hi _A = \Hi_1 \otimes \Hi _3$ 
 and $\Hi _B = \Hi _2
 \otimes \Hi_4$ as local spaces with respect to a LOCC operation $\Lambda$.
Since the local copying protocol is closely related to entanglement
catalysis \cite{ACP, JP}, that is well known open problem,
it is very hard to derive a necessary and sufficient condition for general
settings of local copying.
On the other hand, it is well known that no maximally entangled state
works as entanglement catalysis.
In this paper, to avoid the difficulty of entanglement catalysis, 
we restrict our analysis to the case where all of $\ket{\Psi _j}$
are maximally entangled states 
\footnote{The paper \cite{ACP} showed that
if a copied set $\{ \ket{\Psi _j} \}_{j=0}^{N-1}$ has at least one maximally
entangled state and is locally
copiable, then all of states $\ket{\Psi _j}$ in the copied set must be
maximally entangled.}. 
Therefore, $\ket{\Psi _j} \in \Hi _1 \otimes \Hi _2$ 
can be represented by a unitary operation $U_j \in \Hi _1$ as
\begin{equation} 
\ket{\Psi _j}=(U_j \otimes i) \ket{\Psi _0}.
\end{equation}

As the preparation of our paper, we shortly summarize 
Anselmi, {\it et. al.}'s necessary and sufficient condition of local
copying as follows \cite{ACP}. 
\begin{Theorem} \label{nscondition} 
A set of maximally entangled states $\{ \ket{\Psi _j} \}_{j=0}^{N-1}$ 
is locally copiable, if and only if there exists a unitary operator $A$ on $\Hi _1 
\otimes \Hi _3$ satisfying 
\begin{equation} \label{nscondition1} 
A(T_{jj^{'}} \otimes I)A^{\dagger} 
=e^{i(\theta _j - \theta _{j^{'}})}(T_{jj^{'}} \otimes T_{jj^{'}}), 
\end{equation} 
where $T_{jj^{'}} = U_j U_{j^{'}}^{\dagger}$.
\end{Theorem}
Since each $\ket{\Psi _j}$ must be orthogonal, each $T_{jj^{'}}$ must satisfy 
$\Tr ( T_{jj^{'}} ) = \delta _{jj^{'}}$.
Actually, we can derive the orthogonal conditions from Equation 
(\ref{nscondition1}) by taking trace of them.
In the above theorem, the local copying operation $\Lambda $ is explicitly
represented as a local unitary transformation $A^{13} \otimes A^{*24}$. 
Finally, they solve the equation (\ref{nscondition1}) for all $j, j^{'}$ in the case of $N=2$. 
In this case, 
there is only one independent equation, and  these conditions are reduced to
the condition $A(T \otimes 
I)A^{\dagger} =T \otimes T$, where the phase factor $e^{i\theta _j}$ is 
absorbed  by $T$.
The following theorem is the conclusion of their analysis of Equation 
(\ref{nscondition1}) for $N =2$.
\begin{Theorem} \label{N=2}
There exists a unitary operator $A$ satisfying 
\begin{equation} A(T 
\otimes I) A^{\dagger} = T \otimes T, 
\end{equation} 
if and only if a unitary 
operator $T$ satisfies the following two conditions:
\begin{enumerate} 
\item The 
spectrum of $T$ is the set of power of $M$th roots of unity, where $M$ is a 
factor of $D$.

\item The distinct eigenvalues of $T$ have equal degeneracy.
\end{enumerate}
\end{Theorem}

Here, we should remark the number of maximally entangled states as the resource.
If we allow to use three entangled states as a resource, we can always locally copy any orthogonal set
of maximally entangled states by use of quantum teleportation \cite{teleportation}, 
(For the case when we share two entangled states as resources, see \cite{GKR}.)

\section{local copying of the maximally entangled states in prime-dimensional systems} \label{local copying}
In this paper, we solve Equation (\ref{nscondition1}) and get the necessary 
and sufficient condition  for all $N$ 
in the case of {\it prime} $D$-dimensional local systems.
That is, the form of $T$ is completely determined. 
As a consequence, we show that $D$ is the maximum size of a locally copiable set.

As the starting point of our analysis, we should remark that
Equation (\ref{nscondition1}) simultaneously presents $N^2$ matrix equations, but we may take care of only $N-1$ equations 
$A (T _{j0} \otimes I) A^{\dagger} = e^{i \theta _j - i \theta _0} T _{j0} \otimes T _{j0} \quad (j = 0, \cdots , N-1)$.
This is because by multiplying the $j$ elements of the equation by the Hermitian conjugate of $j^{'}$ elements of 
the same equation, we can recover Equation (\ref{nscondition1}).
Moreover, since $T _{j0}=U_j U_0^{\dagger} = U_j$ and the coefficient $e^{i \theta _j}$ is only related to the unphysical global phase factor, we can treat only the following $N$ equations, 
\begin{equation} \label{fundamental} 
A (U _{j} \otimes I)A^{\dagger} = U _{j} \otimes U _{j} \quad (j = 0, \cdots , N-1).
\end{equation}
Note that $\ket{\Psi _j}$  is represented as $\ket{\Psi _j} = U_j \otimes I 
\ket{\Psi _0}$.

At the first step, we construct an example of a locally copiable set of $D$ maximally entangled states.
\begin{Theorem}\label{sufficiency}
When the set of maximally entangled states $\{ \ket{\Psi _j} \}_{j=0}^{N-1}$ is defined by \begin{equation} 
\ket{\Psi _j} = (U_j \otimes I) \ket{\Psi _j} 
\end{equation} 
and 
\begin{equation} \label{definition} 
U_j = \sum _{j=0}^{D-1} \omega ^{jk} \ket{k} \bra{k}, \end{equation} 
where $\{ \ket{k} \}_{k=0}^{D-1}$ is an orthonormal basis of the $\Hi _1$, then 
the set $\{ \ket{\Psi _j} \}_{j=0}^{N-1}$ can be locally copied.
\end{Theorem}
\begin{Proof}
We define the unitary operator $A$ by 
\begin{equation} \label{repA} 
A = \sum _{a,b,c} \xi _{b,c}^{a} \ket{a \ominus c} \ket{c} \bra{a} \bra{b}, 
\end{equation} 
where $\xi _{b,c}^a$ is a unitary for the indices ${b,c}$ and $\ominus$ means subtraction modulus $D$.
Then, we can easily verify Equation (\ref{fundamental}) as
\begin{eqnarray*} 
& & A(U_j \otimes I)A^{\dagger} \\
& = & \sum \xi _{b_1 c_1}^{a_1} \ket{a_1 \ominus c_1} \ket{c_1} \bra{a_1} \bra{b_1} (\omega ^{j a_3} \ket{a_3} \bra{a_3}) \\
& & \qquad \overline{\xi} _{b_2 c_2}^{a_2} \ket{a_2} \ket{b_2} \bra{a_2 \ominus c_2} \bra{c_2} \\ & = & \sum _{a_1, c_1} \omega ^{ja_1} \ket{a_1 \ominus c_1} \bra{a_1 \ominus c_1} \otimes \ket{c_1} \bra{c_1} \\ & = & \sum _{d, c_1} \omega ^{j d} \ket{d} \bra{d} \otimes \ket{c_1} \bra{c_1} \\ & = & U_j \otimes U _j, 
\end{eqnarray*} 
where we set $d = a_1 \ominus c_1$.
Therefore, Theorem \ref{nscondition} guarantees that the set $\{ \ket{\Psi _j} 
\}_{j = 0}^{D-1}$ can be locally copied.
\hfill $\square$
\end{Proof}
Here, we should remark that $U _1$ is the generalized Pauli's $Z$ operator which is one of 
the generators of the Weyl-Heisenberg Group, and another $U _j$ is the $j$th power of $U_1 = Z$.
Hence, in the case of non-prime-dimensional local systems, 
the spectrum of $U _j$ is different from that of $U _1$
if $j$ is a non-trivial factor of $D$.

Moreover, the property of Weyl-Heisenberg Group not only guarantees that the above example 
satisfies  (\ref{fundamental}), but also is essential for the condition 
(\ref{fundamental}).
That is, as is proved below, any locally copiable set of maximally entangled states 
is restricted exclusively to the above example.
Therefore, our main theorem can be written down as follows. 
\begin{Theorem} \label{main} 
In prime-dimensional local systems, the set of maximally entangled states $\{ U_j \otimes I \ket{\Psi _0} 
\}_{j=0}^{N-1}$ 
can be locally copied if and only if there exist an orthonormal basis $\{ \ket{a }\}_{a=0}^{D-1}$ and 
a set of integers $\{n _j \}_{j=0}^{N-1}$ such that the unitary $U _j$ can be written as 
\begin{equation} \label{nsrep} 
U_j = \sum _{a=0}^{D-1} \omega ^{n_j k} \ket{a} \bra{a}, 
\end{equation} 
where $\omega$ is the $D$th root of unity.
\end{Theorem}
Since the size of the set $\{ U_j \}$ is $D$, $D$ is the maximum size of a locally copiable set of maximally entangled 
states with  prime-dimensional local systems.
\begin{Proof}
(If part)
We have already proven that $\{ U_j \otimes I \ket{\Psi _j} \}_{j=0}^{D-1}$ can be copied by LOCC in 
Theorem \ref{sufficiency}.
Therefore, the subset of them can be trivially copied by LOCC.

(Only if part)
Assume that a unitary operator $A$ satisfies the condition (\ref{fundamental}) for all $j$.

By applying Theorem \ref{N=2}, we can choose an orthonormal basis $\{ \ket{a} \}_{a=0}^{D-1}$ such that 
\begin{equation} \label{expression}
U_1 = \sum _{a=0}^{D-1} \omega ^a \ket{a} \bra{a}, 
\end{equation} 
where $\omega$ is $D$th root of unity.
Moreover, Equation (\ref{fundamental}) implies that the unitary $A$ should transform the subspace $\ket{a} \otimes \Hi$ to subspace $span \{ \ket{k} \otimes \ket{l} \}_{k \oplus l = a}$.
That is, $A$ is expressed as
\begin{equation} \label{repA}
A = \sum _{a,b,c} \xi _{b,c}^{a} \ket{a \ominus c} \ket{c} \bra{a} \bra{b}, 
\end{equation} 
where $\xi _{b,c}^{a}$ is a unitary matrix for $b, c$ for the same $a$, that is, 
$\sum _{c = 0}^{D-1} \xi _{b,c}^{a} \overline{\xi}_{b^{'} ,c}^{a} = \delta _{b,b^{'}} $
 and $\sum _{b = 0}^{D-1} \xi _{b,c}^{a} \overline{\xi}_{b ,c^{'}}^{a} = \delta _{c,c^{'}}$.
Thus, based on the basis $\{ \ket{a} \}_{a=0}^{D-1}$, Equation (\ref{fundamental}) for all 
$a_1, a_2, b_1, b_2$ is written down as 
\begin{equation} \label{elementeq} 
\bra{a_1}\bra{b_1} A(U_j \otimes 1) A^{\dagger} \ket{a_2}\ket{b_2} =\bra{a_1} U_j \ket{a_2} \bra{b_1} U_j \ket{b_2}.
\end{equation}
Therefore, substituting Equation (\ref{repA}) to Equation (\ref{elementeq}) for any integer $j$, we obtain 

\begin{eqnarray} \label{matrixeq} 
\sum _{b=0}^{D-1} \xi _{b,b_1}^{a_1 \oplus b_1} \overline{\xi} _{b, b_2}^{a_2 \oplus b_2}
\bra{a_1 \oplus b_1} U_j \ket{a_2 \oplus b_2} \nonumber \\ = \bra{a_1} U_j \ket{a _2} \bra{b_1}U_j \ket{b_2}, 
\end{eqnarray} 
for all $a _1, a_2, b_1$ and $b_2$.

To see that $U _1 $ and $U _j$ can be simultaneous orthogonalized, we need to prove the following lemma. 
\begin{Lemma} \label{lemma} 
A non-zero $D \times D$ matrix $U_{ab}$ satisfies the following equation, 
\begin{equation} \label{eqlemma1} 
\Xi _{b_1 b_2}^{a_1 \oplus b_1 , a_2 \oplus b_2} U_{a_1 \oplus b_1 \ a_2 \oplus b_2}
=U_{a_1 a_2} U_{b_1 b_2},
\end{equation}
where $\Xi _{b_1 b_2} ^{c c} = \delta _{b_1 b_2}$ and all indices have their value between $0$ and $D - 1$, then $U_{ab}$ is a diagonal matrix.
\end{Lemma}
\begin{Proof}
See Appendix A
\end{Proof}


We apply this Lemma \ref{lemma} to the case when $U_{ab} = \bra{a} U_j \ket{b}$ and $\Xi _{b_1 \ b_2}^{a_1 \ a_2} = 
\sum _{b = 0}^{D-1} \xi _{b b_1}^{a_1} \overline{\xi} _{b b_2}^{a_2}$.
Then, this lemma shows that $U_j$ is orthogonal in the eigenbasis of $U_1$, therefore all 
unitaries $\{U_j \}_{j=0}^{N-1}$ are orthogonalized.
Then, we can get the form of $\{ U_j \}_{j=0}^{N-1}$ explicitly as follows.
From the diagonal element of (\ref{matrixeq}), we derive
\begin{equation} \label{group}
\bra{a \oplus b} U_j \ket{a \oplus b} = \bra{a} U_j \ket{a} \bra{b} U_j \ket{b}.
\end{equation}
Since $\{ \ket{a} \}_{a=0}^{D-1}$ is also an eigenbasis of $U_j$, we can express $U _j $ as 
\begin{equation} \label{unitary} 
U_j = \sum _{a = 0}^{D-1} \omega^{P_j (a)} \ket{a} \bra{a}, 
\end{equation} 
where $P_j (a) $ is a bijection from $\{ a \}_{a=0}^{D-1}$ to themselves.
Then, Equation (\ref{group}) guarantees that $P_j (a)$ is a self-isomorphism of the cyclic group 
$\{ a \}_{a=0}^{D-1}$.
Since a self-isomorphism of a cyclic group is identified by the image of the generator, 
 we derive the formula (\ref{nsrep}) with $P_j(1) = n_j$.
\hfill $\square$
\end{Proof}

We have solved the LOCC copying problem only for a {\it prime}-dimensional local space. 
In the case of a non-prime-dimensional local space, our proof of the ``{\it only if part}'' can be done in the same way. 
However, the ``{\it if part}'' is extended straightly only for the case in which 
the set $\{ U_j \}_{j=0}^{N-1}$ contains at least one unitary whose eigenvalues are generated by
the $D$th root of unity.
In this case, the proof is the following.
By the same procedure of the prime-dimensional case, we obtain Equation (\ref{matrixeq}). 
Then, Lemma (\ref{lemma}) yields all $U_j$ can be orthogonalized and also implies Equation
(\ref{group}) for all $U_j$. By writing $U_j$ as (\ref{unitary}), 
we get the equation $P_j (a \oplus b) = P _j(a) \oplus P _j(b)$ and, so, $P _j(a) = a P _j(1)$. 
Hence, Theorem \ref{N=2} guarantees the same representation of $U_j$ as (\ref{nsrep}). 
Therefore, we can solve the problem of local copying in non-prime-dimensional local spaces only in this special case 
as the direct extension of Theorem \ref{main}. 
On the other hand, if eigenvalues of all $U_j$ are degenerate, 
our proof of ``{\it if part}'' does not hold.

\section{relation with LOCC copying and LOCC discrimination} \label{relation}
If we have no LOCC restriction, the possibility of the deterministic copying is
equivalent with that of the deterministic distinguishability.
Recently, LOCC discrimination problems have been also studied by many researchers
and they found that the LOCC restriction causes the difficulty of 
the discrimination of entangled states
\cite{WSHV00, VSPM01, CY02, C04, F04, VP03, GKRS02}.
Since both copying and discrimination problems are the most essential
problems for characterizing the difference between quantum information theory and classical information theory,
we can derive deeper knowledge about quantum information  through their analysis, especially about entanglement theory.
In this section, we compare the locally distinguishability and the locally copiability for a set of
orthogonal maximally entangled states. 
At first, we remind the definition of a locally distinguishable set, 
and then mention several known and new results of locally distinguishability.
A set of states $\{ \ket{\Psi _j} \}_{j=0}^{N-1}$ is called locally distinguishable, if
there exists a POVM $\{ M_j \}_{j=0}^{N-1}$ which can be constructed by LOCC and also satisfies 
the following conditions:
\begin{equation}
\forall i,j ,\quad \bra{\Psi _i} M_j \ket{\Psi _i} = \delta _{ij}.
\end{equation}
In order to compare LOCC copying and LOCC discrimination, we should take care of the following point:
We assume an extra maximally entangled state 
only in the LOCC copying case.
This is because  
LOCC copying of a set of maximally entangled states is trivially impossible without a blank entangled state. 
This fact is contrary to LOCC discrimination since LOCC discrimination requires
sharing no maximally entangled state.

In the previous section, we already proved that 
$D$ is the maximum size of locally copiable set of maximally entangled states. 
In the case of local discrimination, we can also prove that 
$D$ is the maximum size of a locally distinguishable set of maximally entangled states. 
This statement was proved by the paper \cite{GKRS} only when the set of maximally entangled states
$\{ \ket{\Psi _j} \}_{j = 0}^{N-1}$ consists of canonical form Bell states,
where a canonical form Bell state $\ket{\Psi _{nm}}$ is defined as 
\begin{eqnarray*}
\ket{\Psi _{nm}} & \stackrel{\rm def}{=} & Z^n X^m \otimes I \ket{\Psi _{00}} \\
\ket{\Psi_{00}} & \stackrel{\rm def}{=} & \sum _{k=0}^{d-1} \ket{k}\otimes \ket{k}\\
X & \stackrel{\rm def}{=}& \sum _{k=1}^d \ket{k}\bra{k \oplus 1}.
\end{eqnarray*}
Such a set is a special case of a set of maximally entangled states.
Here, we give a simple proof of this statement for a general set of maximally entangled states.
\begin{Theorem} \label{dika}
If an orthogonal set of maximally entangled states $\{ \ket{\Psi _j} \} _{j=0}^{N-1}$ is locally distinguishable,
then $N \le D$.
\end{Theorem}
\begin{Proof}
Suppose that $\{ M_j \}_{j =0}^{N-1}$ is a separable POVM which distinguishes $\{ \ket{\Psi _j} \}_{j=0}^{N-1}$, 
then they can be decomposed as 
$M _i = \sum _{k=0}^{L} p_{ik} \ket{\psi _k}\bra{\psi _k} \otimes \ket{\phi _k} \bra{\phi _k}$, where $p_{ik}$
is a positive coefficient.
Then, we can derive an upper bound of $\bra{\Psi _j} M_i \ket{\Psi _j}$ as follows,
\begin{eqnarray*}
\bra{\Psi _j} M_i \ket{\Psi _j} & = & \sum _{k=0}^{L} p_{ik} \bra{\Psi _j} \ket{\psi _k}\bra{\psi _k} \otimes 
                                     \ket{\phi _k} \bra{\phi} \ket{\Psi _j} \\
                                & \le & \sum _{k=0}^{L} p_{ik} \bra{\psi _k} ( \frac{1}{D} I) \ket{\psi _k} \\
                                & \le & \frac{\Tr M _i}{D},
\end{eqnarray*}
where the first inequality is come from the montonisity of the fidelity under partial trace operations
concerning the system $B$.
Since $\bra{\Psi _j} M_j \ket{\Psi _j} = 1$, we have $1 \le \Tr ( M _j ) / D$.
Finally, taking the summation of the inequality for $j$, we obtain $N \le D^2/D=D$.
\hfill $\square$
\end{Proof}

When we consider the relationship between local discrimination and local copying of a set of maximally entangled states,
it is quite useful to introduce ``\textit{Simultaneous Schmidt Decomposition}'' \cite{HH04}.
A set of states $\{ \ket{\Psi _{\alpha}} \} _{\alpha \in \Gamma} \subset \Hi _1 \otimes \Hi _2$ is
called simultaneously Schmidt decomposable, if they can be written down as 
\begin{equation} 
\ket{\Psi _{\alpha}} = \sum _{k=0}^{d-1} b_k^{( \alpha )} \ket{e_k} \ket{f_k}, 
\end{equation} 
where $\Gamma$ is a parameter set, $\{ \ket{e_k} \}_{k=0}^{d-1}$ and $\{ \ket{f_k} \}_{k=0}^{d-1}$ are 
orthonormal bases of local spaces
(simultaneous Schmidt basis) and $b_k^{(\alpha )}$ is a {\itshape complex number} coefficient. 
Actually, simultaneous Schmidt decomposability is a sufficient condition for local distinguishability
of a set of maximally entangled states
and a necessary condition for local copiability of it.
Moreover, simultaneous Schmidt decomposability is not a 
necessary and sufficient condition for the both cases. 
Therefore, a family of locally copiable sets of maximally entangled states is strictly included by 
a family of locally distinguishable sets of maximally entangled states.
In the following, we prove this relationship.

First, we explain the relationship between local discrimination and simultaneous Schmidt decomposition which has
been already obtained by the paper \cite{VSPM01}. 
Any unknown state $\ket{\Psi _{\alpha}} \in \Hi _A \otimes \Hi _B$ in
a simultaneously Schmidt decomposable set of states $\{ \ket{ \Psi _{\alpha}} \} _{\alpha \in \Gamma}$ can be
transformed to a single local space $\Hi _A$ or $\Hi _B$ by LOCC.
Rigorously speaking, there exists a LOCC $\Lambda$ on $\Hi _A \otimes \Hi _{B_1 B_2}$ which transforms 
$\ket{\Psi _{\alpha}^{AB_1}} \otimes \ket{0^{B_2}}$ to $\sigma^A \otimes \ket{\Psi _{\alpha}^{B_1 B_2}}$ 
for all $\alpha \in \Gamma$, and also exists a LOCC $\Lambda ^{'}$on $\Hi _{A_1} \otimes \Hi _{A_2} \otimes \Hi _B$ 
which transforms 
$\ket{0^{A_1}}  \otimes \ket{\Psi _{\alpha}^{A_2B}} $ to $\ket{\Psi _{\alpha}^{A_1 A_2}} \otimes \sigma^B  $ 
for all $\alpha \in \Gamma$.
Indeed, this LOCC transformation can be written down as the following Kraus representation \cite{VSPM01}:
\begin{eqnarray*}
&\rho \mapsto 
\sum _{k=0}^{d-1} F_k
\rho 
F_k^*,
\end{eqnarray*}
where
\begin{eqnarray*}
F_k
  & \stackrel{\rm def}{=} &
(I_A \otimes CNOT)
(U_k \otimes I_{A,B_2})
(P_k \otimes I_{B_1,B_2}) \\
P_k & \stackrel{\rm def}{=} & 1/D(\sum _{i } \omega ^{ki} \ket{e_i})(\sum _{l } \omega ^{kl} \bra{e_l}) \\
U_k & \stackrel{\rm def}{=} & \sum _{i} \omega ^{ki} \ket{f_i} \bra{f_i} \\
CNOT & \stackrel{\rm def}{=} & \sum _{kl} \ket{e_k} \otimes \ket{f_{k \oplus l}} \bra{f_k} \otimes \bra{l}.
\end{eqnarray*}
In the above formula, both $\{ \ket{e_k} \}_{k=0}^{D-1}$ and $\{ \ket{f_l} \}_{k=0}^{D-1}$ are 
simultaneous Schmidt bases of $\{ \ket{\Psi _{\alpha}} \}_{\alpha \in {\Gamma}}$,
and $\{ \ket{l} \}_{l=0}^{D-1}$ is the standard computational basis.
Therefore, 
if a set $\{ \ket{\Psi _{\alpha} }\}$ is
simultaneously Schmidt decomposable,
there exists a one-way-LOCC POVM $M'=\{M'_i\}$ 
for a given arbitrary POVM $M=\{M_i\}$ such that
\begin{eqnarray*}
\bra{\Psi _{\alpha}}  M_i \ket{\Psi _{\alpha}} 
=
\bra{\Psi _{\alpha}}  M'_i \ket{\Psi _{\alpha}} ,\quad
\forall i, \forall \alpha.
\end{eqnarray*}
That is, any POVM can be essentially realized by one-way LOCC.
Therefore, ``\textit{simultaneously Schmidt decomposable set of orthogonal maximally entangled states 
is locally distinguishable.}'' On the other hand, the set of orthogonal maximally entangled states
which is not simultaneously Schmidt decomposable was found by the paper \cite{GKRS}. 
Thus, a family of simultaneously Schmidt decomposable sets of maximally entangled states is strictly included by 
a family of locally distinguishable sets of maximally entangled states.  

On the other hand, the relationship between simultaneous Schmidt decomposability and
local copiability can be derived as the following theorem. 
\begin{Theorem}
In prime-dimensional local systems, an orthogonal set of maximally entangled states $\{ \ket{\Psi _j} \}_{j=0}^{N-1}$
is locally copiable,  
if and only if it is a  simultaneously Schmidt decomposable 
subset of canonical form Bell states under the same local unitary operation.
\end{Theorem}
\begin{Proof}
We can easily see the ``{\itshape only if}'' part of the above statement from Theorem \ref{main}.
The ``{\itshape if}'' part can be showed as follows.
The paper \cite{HH04} shows that the states $\ket{\Psi _{n_{\alpha} m_{\alpha}}} (\alpha = 1,2, \cdots , l)$ 
are simultaneously Schmidt decomposable, if and only if there exist integers $p$, $q$ and $r$ 
($p \neq 0$ or $q \neq 0$) satisfying $p n_{\alpha} \oplus q m_{\alpha} = r$ for all $\alpha$. 
Since the ring ${\mathbb Z}_p$ is a field in the prime number $p$ case, the
above condition is reduced to the existence of $f$ and $g$ such that $m_{\alpha} =f n_{\alpha} + g$.
Then, we get
\begin{eqnarray}
\ket{\Psi _{n_{\alpha} m_{\alpha}}} & = & \ket{\Psi _{n_{\alpha} (f n_{\alpha} +g)}} \nonumber \\
                                    & = & C_{\alpha} (Z X^f)^{n_{\alpha}} X^g \otimes I \ket{\Psi _{00}}. \label{eqssd} 
\end{eqnarray} 
Since $Z X^f$ is unitary equivalent to $Z$ \cite{F04}, the state $\ket{\Psi _{n_{\alpha} m_{\alpha}}} $ is locally unitary 
equivalent with $U_j \otimes I \ket{\Psi _0}$ in Theorem \ref{main}.
\hfill $\square$
\end{Proof}
Therefore, a family of locally copiable sets of maximally entangled states is strictly included by
a family of simultaneously Schmidt decomposable sets of maximally entangled states. 
In other words, local copying is much more difficult than local discrimination.

\section{summary} \label{summary}

Based on the analysis by Anselmi {\it et. al.} \cite{ACP}, we completely characterized LOCC copying of orthogonal 
maximally entangled states in prime-dimensional local systems. 
As our result, it was found that the assumption of LOCC not only reduces the maximum size of a copiable set of 
maximally entangled states to $D$, 
but completely determines the form of states with the equivalence of local unitary transformations (Theorem \ref{main}).
In the next step, to compare the local copiability with the local distinguishability,
we gave a simple proof that the maximum number of local distinguishable maximally entangled states
is also $D$.
Then, we showed that in prime-dimensional local systems, LOCC copying of a set of maximally entangled states 
is possible, if and only if it is a simultaneously Schmidt decomposable subset 
of canonical form generalized Bell states under the same local unitary transformation.
As a conclusion,
since a simultaneously Schmidt decomposable set is locally distinguishable \cite{VSPM01}, 
LOCC copying is more difficult than  LOCC discrimination in prime-dimensional local systems
for orthogonal maximally entangled states.

Finally, we should mention a remained open problem.
In this paper, we showed the necessity of the form of states (\ref{nsrep}) for LOCC copying only in prime-dimensional local systems.
However, we restrict this dimensionality only by the technical reason, and
this restriction has no physical meaning. Thus, the validity of Theorem
\ref{main} for non-prime-dimensional systems still remains as a open question.

After finishing this research, the authors found a related paper \cite{Nathanson} which contains
a different approach to Theorem \ref{dika}.

\section*{Acknowledgments}
The authors would like to express our gratitude to Mr. F. Anselmi, Dr. A. Chefles and Professor M.B. Plenio for their useful discussion.
They would like to thank S. Virmani for useful discussion.
They are particularly indebted to Professor M. Murao for her helpful advice and discussion.
They are grateful to Professor K. Matsumoto for his advice, Dr. T. Hiroshima for helpful comments on the introduction and 
Professor H. Imai for his support and encouragement.
They are also grateful Professor M.B. Ruskai for informing them the paper \cite{Nathanson}.


\appendix
\section{Proof of Lemma \ref{lemma}}
First, by choosing $c = a_1 \oplus b_1 = a_2 \oplus b_2$, we have 
\begin{equation} \delta _{b_1 b_2} U_{cc} = U_{c \ominus b_1 \ c \ominus b_2} U_{b_1 b_2}.
\end{equation}
In addition, choosing $b_1 \neq b_2$, we derive 
\begin{equation} 
U_{c \ominus b_1 \ c \ominus b_2} U_{b_1 b_2} =0 
\end{equation} 
for all $c$.
The above equation means,
\begin{equation} \label{eqlemma2}
 b_1 \neq b_2 \quad \Longrightarrow \quad U_{b_1 b_2} = 0 \quad or \quad \forall c, \  U_{c \ominus b_1 \ c \ominus b_2} = 0.
\end{equation}

By means of the above fact, we prove that $U_{b \ b \oplus n} = U_{b \ b \ominus n}=0$ for all $b$ by 
induction concerning the integer $n$.
At first, we prove  $U_{b \ b \ominus 1} = 0$ for all $b$ by a contradiction.
We assume there exists $b_1$ such that $U_{b_1 \ b_1 \ominus 1} \neq 0$, 
then, Equation (\ref{eqlemma2}) implies  $U_{b \ b+1} = 0$ for all $b$.
On the other hand, Equation (\ref{eqlemma1}) guarantees that 
\begin{equation} \label{keylemma} 
\Xi _{b_1 \ b_1 \ominus 1}^{b \ b+1} U_{b \ b+1} = U_{b \ominus b_1 \ b \ominus b_1 \oplus 2} U_{b_1 \ b_1 \ominus 1}. 
\end{equation}
Therefore,  $U_{b \ b \oplus 2} = 0$ for all $b$.
Thus, repeatly using (\ref{eqlemma1}), we have $U_{b \ b \oplus a} = 0$ for all $a$ and $b$.
This is a contradiction for the fact that $U_{ab}$ is a non-zero matrix.
So, we have $U_{b \ b \ominus 1} = 0$ for all $b$.
Similarly, we can prove $U_{b \ b \oplus 1} =0$ for all $b$ by a contradiction as follows.
Suppose there exists $b_1$ such that $U_{b_1 \ b_1 \oplus 1}=0$, then Equation (\ref{eqlemma1}) implies $\Xi _{b_1 \ b_1 \ominus 1}^{b \ b \ominus 1} U_{b \ b \ominus 1} = U_{b \ominus b_1 \ b \ominus (b_1 - 2)} U_{b_1 \ b_1 \oplus 1}$.
Therefore, $U_{b \ b \ominus 2} =0$ for all $b$, and repeating this procedure, we have $U_{b \ b \ominus a} =0$
for all $a$ and $b$.
This is a contradiction.
Therefore,  $U_{b \ b \oplus 1} = 0$ for all $b$.

At the next step, we assume  $U_{b \ b \ominus k} = U_{b \ b \oplus k} = 0$ for all $k \le n-1$ and show 
$U_{b \ b \ominus n} = U_{b \ b \oplus n} = 0$ for any $b$ by a contradiction.
Assume that there exists $b_1$ such that  $U_{b_1 \ b_1 \ominus n} \neq 0$, then Equation (\ref{keylemma}) implies that $U _{b \ b \oplus n} = 0$ for all $b$.
Thus, Equation (\ref{eqlemma1}) implies
\begin{equation}
\Xi _{b_1 \ b_1 \ominus 1}^{b \ b \oplus k} U_{b \ b \oplus k} = U_{b \ominus b_1 \ b \ominus b_1 \oplus (n + k)} U_{b_1 \ b_1 \ominus n}.
\end{equation}
Thus, we have $U_{b \ b \oplus (n+k)} =0$ for all $k \le n-1$ and $b$.
Repeating this procedure, we have $U_{b \ b \oplus a} = 0$ for all $a$ and $b$.
This is a contradiction.
Therefore, we have $U_{b \ b \ominus n} = 0$ for all $b$.
Similarly, we can prove $U_{b \ b \oplus n} = 0$ for all $b$.
Finally, by the mathematical induction, we prove $U_{b \ b \oplus n} = U _{b \ b \ominus n} = 0$
for all $0 \le n < D$ and  $b$.
Therefore, $U_{ab}$ is a diagonal matrix.
\hfill $\square$

\end{document}